\documentclass[aps, preprint]{revtex4}
\begin{document}
\title{Comments on "Geometric Phase Associated with Transformations of Optical
Beams Bearing Orbital Angular Momentum"}
\author{S.C. Tiwari}
\affiliation{Institute of Natural Philosophy, \\ 1, Kusum Kutir, Mahamanapuri,
Varanasi - 221 005} 
\maketitle
The first direct measurement of the geometric phase acquired by the  light
beams carrying orbital angular momentum (OAM) has been reported by Galvez et
al \cite{1}. The closed path traversed by the optical beams is the space of
modes of Laguerre-Gaussian beams \cite{2}. Modes $LG^l_p$ carry OAM of $l\hbar$
per photon, where $l$ is the azimuthal index, and $p$ is the radial index. In
the experiments, a geometric phase on a $LG^{-1}_0$ beam is introduced,
changed, and interfered with a reference $LG^0_0$ beam. Authors conclude the
paper with the statements: "All of these geometric phases arose from
transformations involving a change in the OAM of the modes. They support the
conjecture that geometric phase arises from the exchange of angular momentum
between the light and optical system [13]". Ref.13 in \cite{1} is an
interesting work by van Enk,here after referred at \cite{3}. In the present
note I make following comments: 
\begin{itemize}
\item[A1.] 
The connection of angular momentum to the geometrical phase in optics was
vaguely hinted at in \cite{4}. In 1992 plausible arguments were presented to
propose that the physical mechanism of geometrical phase in optics was the transfer
of angular momentum between the light and optical elements \cite{5}. This
conjecture was theoretically supported by van Enk \cite{3} ,and later by
Banerjee \cite{6}. In \cite{3} van Enk considered LG beams for the geometric
phases. 
\item[A2.]
In 1994, LeFloch showed interest to verify this conjecture in his
laboratory, however the extreme difficulty in the measurement of angular
momentum transfer was pointed out by him \cite{7}. The present work measures
geometric phase, and is significant; however the direct demonstration of
angular momentum exchange remains elusive. I believe a direct experimental
verification of this relationship, and measurement of angular momentum exchange would have far reaching implications on the nature of light and
crystal optics \cite{8}. 
\end{itemize}

{\it Acknowledgement}

The library facility of Banaras Hindu University, Varanasi is acknowledged.

\end{document}